\documentclass[sigconf]{acmart}
\usepackage[cjk]{kotex}

\usepackage{soul,color}
\usepackage{algorithm}
\usepackage[noend]{algpseudocode}

\algdef{SE}[DOWHILE]{Do}{doWhile}{\algorithmicdo}[1]{\algorithmicwhile\ #1}%
\usepackage{multirow}
\usepackage{hhline}
\usepackage{array}
\usepackage{amsmath}
\usepackage{setspace}
\usepackage[capitalise]{cleveref}
\usepackage{xcolor}

\usepackage{caption}
\usepackage{makecell}
\usepackage{tabularx}

\usepackage{booktabs}
\usepackage{amsmath}
\usepackage{array}

\copyrightyear{2027}
\acmYear{2027}
\setcopyright{cc}
\setcctype{by}
\acmConference[ASPDAC '27]{32nd Asia and South Pacific Design Automation Conference}{January 25--28, 2027}{Tokyo, Japan}
\acmBooktitle{32nd Asia and South Pacific Design Automation Conference (ASPDAC '27), January 25--28, 2027, Tokyo, Japan}
\acmDOI{10.1145/3840404.3848262}
\acmISBN{979-8-4007-2923-2/2027/01}

\AtBeginDocument{%
  \providecommand\BibTeX{{%
    \normalfont B\kern-0.5em{\scshape i\kern-0.25em b}\kern-0.8em\TeX}}}

\begin{document}
\title{Clock Distribution Network with Backside Clock Mesh\\in BSPDN Architecture}

\author{Seohyun Kim}
\affiliation{%
  \institution{Korea Advanced Institute of Science and Technology}
  \city{Daejeon}
  \country{Republic of Korea}}
\email{shkim1122@kaist.ac.kr}

\author{Youngjin Chae}
\affiliation{%
  \institution{Korea Advanced Institute of Science and Technology (KAIST)}
  \city{Daejeon}
  \country{Republic of Korea}}
\email{yjchae@kaist.ac.kr}

\author{Shilong Zhang}
\affiliation{%
  \institution{Korea Advanced Institute of Science and Technology}
  \city{Daejeon}
  \country{Republic of Korea}}
\email{zhangshilong@kaist.ac.kr}

\author{Youngsoo Shin}
\affiliation{%
  \institution{Korea Advanced Institute of Science and Technology}
  \city{Daejeon}
  \country{Republic of Korea}}
\email{youngsoo@kaist.edu}

\thanks{This work was supported by Samsung Electronics Co., Ltd (IO250219-12006-01) and National Research Foundation of Korea (NRF) grant funded by the Korea government (MSIT) (No.~RS-2024-00405495, Plug\&Play (P\&P) chiplet integration research center).}

\begin{abstract}
We propose a clock distribution network (CDN) in which the clock mesh is located on backside metal (BM) layers, while the rest of the network remains on the frontside. Since the backside power distribution network (BSPDN) also uses the BM layers, the clock mesh and power network must share them. We identify an important problem of placing mesh receivers while accounting for keep-out zones (KOZs) between clock- and power-nTSVs. We show that this problem can be solved using a potential-field method (PFM).

Machine learning models are developed to place the mesh receivers: a U-Net predicts their initial positions, and an MLP predicts force weights for PFM-guided position refinement. Both models are trained through self-supervised learning by optimizing a differentiable objective that accounts for clock latency, IR-drop, and KOZ violations. Experimental results show that, on average, the proposed CDN reduces maximum local clock latency by 26\%, global clock skew by 40\%, and mesh switching power by 29\% compared with a frontside CDN \cite{tap_driver}. On the same backside CDN, ML-guided PFM satisfies KOZ constraints while lowering IR-drop by 33\% and requiring 85\% fewer iterations than standard PFM.

\end{abstract}


\maketitle

\section{Introduction}
A clock mesh is popular in high-performance designs thanks to smaller clock skew \cite{sitik2012multi} and less susceptibility to on-chip variations \cite{li2025clock}. The challenges of mesh design include (1) large switching power consumption \cite{abdelhadi2013timing} due to higher load capacitance of mesh wires, (2) short-circuit power \cite{shim2013analysis} consumed by mesh drivers, and (3) degraded signal routability due to the metal resources occupied by the mesh.

\begin{figure}[t]
  \centering \includegraphics[scale = 1.0]{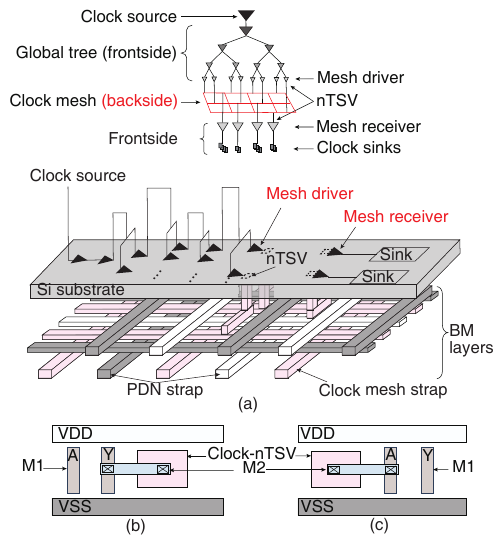}
  \caption{(a) Proposed CDN with backside clock mesh, (b) mesh driver, and (c) mesh receiver.}
  \vspace{-4.5mm}
  \label{fig:bsmesh}
\end{figure}

As technology node shrinks, the wafer backside is being utilized for power distribution network (PDN) \cite{bspdn} or for clock tree routing \cite{bsclock} to help signal routability on the frontside. Backside metal (BM) layers can be good candidates for clock mesh since they are thicker than frontside metals by 2-3 times \cite{thick}.
Their resistance, therefore, is lower. Their capacitance is also lower by 20\%; larger spacing reduces coupling capacitance and the increased thickness of BM places the wires farther from the substrate, reducing ground capacitance.

In this paper, we propose a new CDN architecture in which the clock mesh is placed on the backside while the rest remains on the frontside, as shown in \Cref{fig:bsmesh}(a). Because the backside power distribution network (BSPDN) also uses the BM layers, the clock mesh and power network now share them. We address the challenge of placing mesh receivers while accounting for keep-out zones (KOZs) between clock- and power-nTSVs.

\subsection{Proposed CDN with Backside Mesh}
As shown in \Cref{fig:bsmesh}(a), the global clock tree is routed on the frontside and connected to the backside clock mesh via clock-nTSVs. Clock signal is returned to the frontside and is connected to mesh receivers, which drive clock sinks.
A mesh driver and mesh receiver (i.e., tap driver) are designed such that clock-nTSV can be landed \cite{bsclock} as shown in \Cref{fig:bsmesh}(b) and (c).

\begin{figure}[t]
  \centering \includegraphics[scale = 0.95]{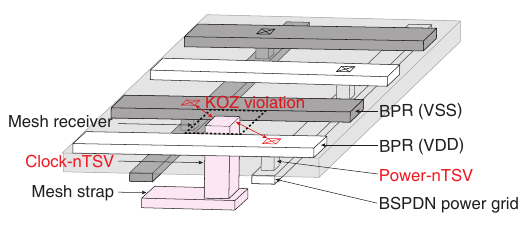}
  \caption{BM layers for clock mesh and PDN, indicating potential KOZ violations between power- and clock-nTSVs.}
  \label{fig:bs_sharing}
\end{figure}

In BSPDN, power grid connects to BPRs (buried power rails) at all intersections via power-nTSVs (see \Cref{fig:bs_sharing}).
If a mesh receiver is introduced, clock-nTSV is required to provide a connection to mesh strap. The constraint of KOZ between power- and clock-nTSV is important, so if some power-nTSVs are too close to particular clock-nTSV, they should be dropped. This in turn can lead to increased IR-drop. Thus, mesh receiver placement should avoid existing IR-drop hotspots to prevent further degradation.

To the best of our knowledge, no previous work directly addresses mesh receiver placement for backside clock mesh. Existing method \cite{tap_driver}, developed for frontside meshes, assumes uniform grids over the floorplan and places one mesh receiver at each grid center (see \Cref{fig:tap_plc}(a)). If it is applied to a backside mesh, KOZ violations and IR-drop cannot be resolved as we will show in Section~4.

\subsection{Problem Definition and Proposed Solution}
\begin{figure}[t]
  \centering \includegraphics[scale = 0.95]{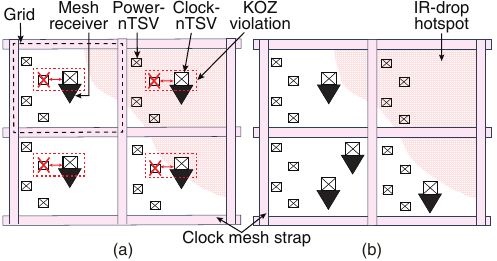}
  \caption{Mesh receivers placed through (a) the method in \cite{tap_driver} and (b) the proposed method. Red boxes are KOZ violations.}
  \vspace{-4mm}
  \label{fig:tap_plc}
\end{figure}

Mesh drivers are assumed to be fixed at all mesh strap intersections \cite{chen2016comparative, sitik2012multi}. Mesh receivers are placed such that max local clock latency is minimized while (1) existing IR-drop hotspots are avoided and (2) all KOZ constraints are respected (\Cref{fig:tap_plc}). Each mesh receiver delivers the clock signal to nearby sinks through a local clock tree.

This problem can be addressed using a potential-field method (PFM) \cite{pfm}, where each mesh receiver is modeled as a particle subject to an attractive force toward its assigned sinks to minimize latency, and repulsive forces from KOZ and IR-drop hotspots to satisfy the placement constraints. The net force is given by
\begin{equation}
\mathbf{F}_\text{net} = \alpha\mathbf{F}_\text{sink}
+ \beta\mathbf{F}_\text{IR} + \gamma\mathbf{F}_\text{KOZ}.
\end{equation}

The standard PFM \cite{pfm}, however, depends heavily on initial position and requires manual tuning of the weights ($\alpha, \beta, \gamma$), which is not trivial. More importantly, the optimal ($\alpha, \beta, \gamma$) depends on the local sink distribution, nTSV density, and IR-drop at each receiver, yet the standard formulation uses fixed, globally shared weights. For example, in regions where sinks are concentrated, IR-drop hotspots are highly likely to occur as well, and $\mathbf{F}_\text{IR}$ strongly repels the receiver; in such regions, $\alpha$ must be increased to ensure the receiver is still attracted close enough to the sinks. Manual assignment of such position-dependent weights to lots of receivers is infeasible.

We propose a 2-stage ML-guided approach to address these limitations.
A U-Net model is employed to predict the initial placement of mesh receivers, replacing random initial placement, and a multilayer perceptron (MLP) predicts position-dependent ($\alpha, \beta, \gamma$) that adapt to local conditions during the force-driven refinement. Both models are trained through self-supervised learning, eliminating the need for ground truth.


\subsection{Design Flow}

After the backside PDN is synthesized and placement step is complete, IR-drop analysis is performed to identify hotspots. The clock mesh is then constructed on BM layers, followed by global clock tree synthesis. The proposed mesh receiver placement is now performed: a U-Net generates the initial placement (presented in \Cref{sec2}), which is refined through ML-guided PFM (covered in \Cref{sec3}). Each mesh receiver is connected to sinks through its local clock tree. Signal routing will finally be performed.

\section{Initial Placement of Mesh Receivers}\label{sec2}

The initial placement is performed through U-Net model, i.e. the model receives four input maps and produces an output map indicating mesh receiver locations. The model is a standard U-Net with four encoder blocks, a 256-channel bottleneck, and four decoder blocks with skip connections.
We employ U-Net because mesh-receiver placement depends on information at two scales: the global sink and IR-drop distributions across the floorplan, and the local geometry around each nTSV to determine KOZ violations. The encoder aggregates the global distributions, while the skip connections preserve the local nTSV geometry.

\subsection{U-Net Input and Output Maps}\label{sec2.1}
The chip layout after global clock tree synthesis is represented as a set of 2D maps, where each pixel corresponds to a placement site. The size of a placement site is given by the PDK (21~nm $\times$ 144~nm in our case).
The U-Net takes four input maps (sink, power-nTSV, mesh driver, and IR-drop) and produces one output map for mesh receiver placement. The sink map is binary: a pixel is set to 1 if the clock pin of a sink is located within the placement site, and 0 otherwise.
The clock pin location is used rather than the cell origin
because it is the actual connection point to the mesh receiver.
The mesh driver and power-nTSV maps are binary: they mark the clock-nTSV locations of mesh drivers (placed at mesh strap intersections) and the power-nTSV locations (placed at BSPDN-power rail intersections), respectively. The IR-drop map is continuous: each pixel overlapping a cell takes the cell's IR-drop value, and unoccupied pixels are 0.

The output map is binary, representing the predicted clock-nTSV locations of mesh receivers. When a pixel is predicted as 1, a clock-nTSV is placed at the corresponding site. All maps use the lower-left corner as the reference point so that pixel distances directly map to the physical distances relevant to KOZ constraints.

The U-Net predicts 320 $\times$ 1280 pixels per inference. For full-chip prediction, only the central 256 $\times$ 1024 pixels are taken and stitched, discarding boundary pixels to avoid inconsistency.

\subsection{Self-Supervised Learning}\label{sec2.2}
Since no ground-truth of mesh receiver placement exists, we train the U-Net with self-supervised loss calculated directly from the U-Net input and output. This requires that the loss function is differentiable over every operation from the U-Net output.

The loss consists of three terms that represent IR-drop penalty, KOZ penalty, and sink-to-receiver distance:
\begin{equation}
  \mathcal{L} = \mathcal{L}_{\mathrm{IR}} + \mathcal{L}_{\mathrm{KOZ}}+\mathcal{L}_{\mathrm{sink}}.
\label{eq:loss}
\end{equation}
This simple sum is sufficient to satisfy all constraints in our
experiments (\Cref{tab:power_perf}).

\subsubsection{Mesh Receiver Selection}
All three loss terms depend on $p_r$, the probability that pixel $r$ is selected as a receiver location. To compute $p_r$, the U-Net's per-pixel output $\ell_r$ is passed through a shifted sigmoid. The sigmoid's threshold is chosen so that approximately $k$ pixels have $p_r \approx 1$, matching the number of receivers to be selected:
\begin{equation}
  p_r = \sigma\!\left(\frac{\ell_r - t}{\tau_p}\right),
\end{equation}
where $\sigma(\,\,)$ is the sigmoid function, $\tau_p$ is a temperature parameter that controls how sharply $p_r$ changes between 0 and 1, and $t$ is the threshold.

The number of selected receivers $k$ is determined by the fanout $F_{th}$ that a single receiver can drive:
\begin{equation}
  k = \left\lceil \frac{N_{\text{sink}}}{F_{\text{th}}} \right\rceil,
  \quad
  F_{\text{th}} = \left\lfloor
    \frac{\alpha_{\mathrm{m}} \cdot C_{\text{max}}}{C_{\text{pin,avg}}}
  \right\rfloor,
\end{equation}
where $C_{\text{max}}$ is the maximum output load capacitance of the receiver, $C_{\text{pin,avg}}$ is the average clock pin capacitance of the sinks (both obtained from the liberty files), and $\alpha_{\mathrm{m}} = 0.5$ is a margin that reserves capacitance for wire load.

\subsubsection{IR-drop Penalty}
This term is the sum of IR-drop values at the selected
receiver pixels:
\begin{equation}
  \mathcal{L}_{\text{IR}} = \sum_r p_r \cdot \text{IR}(r).
\end{equation}
Minimizing $\mathcal{L}_{\mathrm{IR}}$ drives receivers toward
low-IR-drop regions.

\subsubsection{KOZ Penalty}
This loss term prevents mesh receiver clock-nTSVs from being placed too close to nTSVs that are already placed (i.e., power-nTSVs and mesh driver clock-nTSVs) or to each other.
Let $v(d) = \max(0,\, d_{\text{KOZ}} - d)^2$ denote the spacing
violation at distance $d$, where $d_{\text{KOZ}}$ is the KOZ distance. The penalty is given by:
\begin{equation}
  \mathcal{L}_{\text{KOZ}} = \sum_{r,f} p_r \, v(d(r,f))
                          + \frac{1}{2}\sum_{r \ne r'} p_r p_{r'} \, v(d(r,r')).
\end{equation}
The first sum penalizes proximity between candidate receivers and
fixed nTSVs; the second penalizes candidate receivers being too close
to each other. The weights $p_r$ (and $p_{r'}$) ensure that only
pixels likely to be selected as receivers contribute.

\subsubsection{Sink-to-Receiver Distance}
\begin{figure}[t]
  \centering \includegraphics[scale = 0.95]{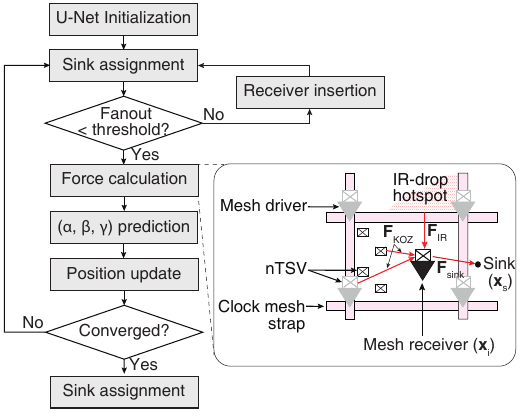}
  \caption{ML-guided PFM for mesh receiver placement.}
  \label{fig:algorithm}
\end{figure}
This term penalizes large sink-to-receiver distances; longer distances increase local clock latency. Each sink $s$ is driven by its nearest receiver $r$, and the maximum local latency is determined by the farthest sink-receiver pair. Computing this involves two non-differentiable operations: (1) assigning each sink to its nearest receiver ($\min_r d(s,r)$), and (2) taking the maximum over all sinks. We introduce a differentiable relaxation for each.

\textbf{Nearest Receiver Distance:}
The hard assignment $d_s = \min_r d(s,r)$ assigns weight 1.0 to the nearest receiver and 0.0 to the rest, zeroing out their gradients. We relax it into a weighted sum \cite{jang2017categorical} over all receivers:
\begin{equation}
\tilde{d}_s = \sum_r w_{sr}\, d(s,r),
\quad
w_{sr} = \frac{p_r \exp\!\left(-d(s,r)/\tau_a\right)}{\sum_{r'} p_{r'} \exp\!\left(-d(s,r')/\tau_a\right)}
\end{equation}
where $w_{sr}$ is larger for receivers that are closer and more likely to be selected, and $\tau_a$ is a temperature controlling assignment softness.

\textbf{Maximum Distance:}
We approximate the maximum over sinks with log-sum-exp:
\begin{equation}
  \mathcal{L}_{\text{sink}} = \tau_l \log \sum_s \exp\!\left(\frac{\tilde{d}_s}{\tau_l}\right)
  \label{eq:L_sink}
\end{equation}
where $\tau_l$ is the temperature controlling max approximation.
We set $\tau_p = 0.02$ and $\tau_a = \tau_l = 0.5\,\mu$m.

\section{Refinement of Mesh Receiver Positions through PFM}\label{sec3}

After the U-Net generates an initial mesh receiver placement, each predicted pixel location is converted to the physical coordinate of its clock-nTSV. Each receiver is then modeled as a virtual particle and its position is refined by ML-guided PFM.

\subsection{Placement Refinement Flow} \label{sec3.1}
As shown in \Cref{fig:algorithm}, sinks are first assigned to their nearest receiver. If any receiver's fanout exceeds the threshold $F_{\text{th}}$ in Eq.~(4), additional receivers are inserted outside the KOZ and sinks are re-assigned; this repeats until no receiver exceeds the threshold.

As illustrated in the inset of \Cref{fig:algorithm}, three forces act on each receiver~$i$ located at~$\mathbf{x}_i$.
Each of its assigned sinks at~$\mathbf{x}_s$ exerts an
attractive force, and the combined force on receiver~$i$ is:
\begin{equation}
  \mathbf{F}_{\mathrm{sink},i}
    = \sum_{s}(\mathbf{x}_s - \mathbf{x}_i).
\end{equation}
Two repulsive forces keep receivers away from constrained regions.
The IR-drop force moves each receiver toward lower IR-drop regions through
\begin{equation}
  \mathbf{F}_{\mathrm{IR},i} = -\nabla\,\mathrm{IR}(\mathbf{x}_i)
\end{equation}
where this gradient is computed on a Gaussian-smoothed IR-drop map.
The KOZ force repels a receiver when it approaches within a KOZ distance~$d_{\mathrm{KOZ}}$ of any nTSV:
\begin{equation}
  \mathbf{F}_{\mathrm{KOZ},i}
    = \max\!\Bigl(0,\;\frac{1}{d}-\frac{1}{d_{\mathrm{KOZ}}}\Bigr)
      \,\hat{\mathbf{u}},
\end{equation}
where $d$ is the center-to-center distance between the receiver clock-nTSV and the nearest conflicting nTSV, and $\hat{\mathbf{u}}$ is the direction in which the KOZ repels the receiver.

The relative importance of each force depends on the local context.
A shared MLP predicts coefficients $(\alpha_i,\beta_i,\gamma_i)$ for each mesh receiver from its local context around~$\mathbf{x}_i$, as detailed in \Cref{sec3.2}.
The net force is then
\begin{equation}
  \mathbf{F}_{\mathrm{net},i}
    = \alpha_i\,\mathbf{F}_{\mathrm{sink},i}
    + \beta_i\,\mathbf{F}_{\mathrm{IR},i}
    + \gamma_i\,\mathbf{F}_{\mathrm{KOZ},i},
  \label{eq:fnet}
\end{equation}
and all mesh receiver positions are updated simultaneously as
\begin{equation}
  \mathbf{x}_i \leftarrow \mathbf{x}_i
    + \eta\,\mathbf{F}_{\mathrm{net},i},
  \label{eq:update}
\end{equation}
where $\eta$ is the step size.
The loop of force calculation, coefficient prediction, and position update repeats until no receiver moves more than one placement site.
After the loop terminates, sinks are assigned to their nearest receiver, and all receiver positions are legalized to the nearest valid placement site.

\subsection{MLP Training}\label{sec3.2}
\begin{figure}[t]
  \centering \includegraphics[scale = 0.95]{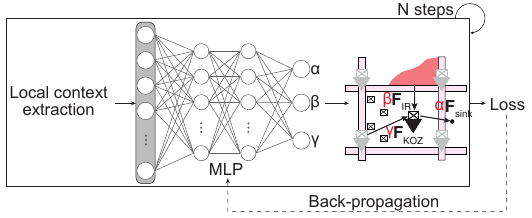}
  \caption{MLP for coefficient prediction.}
  \vspace{-2mm}
  \label{fig:mlp}
\end{figure}
The MLP predicts coefficients for PFM refinement. As input, it takes five local feature maps cropped from a $3d_{\mathrm{KOZ}} \times 3d_{\mathrm{KOZ}}$ region centered at~$\mathbf{x}_i$. Four of these are the U-Net input maps (sink map, power-nTSV map, mesh driver map, and IR-drop map; see \Cref{sec2.1}), which remain static during refinement. The fifth is a density map of neighboring receivers, where each receiver $j \neq i$ is represented as a Gaussian kernel centered at $\mathbf{x}_j$ to ensure differentiability. All five maps are flattened and concatenated into a single context vector.

The context vector is passed through a three-layer MLP (725--64--64--3) with ReLU activations, which outputs positive coefficients via a sigmoid function. The coefficients are kept positive because the force directions are fixed by the formulation in \Cref{sec3.1}; the MLP controls only their magnitudes.

The MLP is trained end-to-end with the same self-supervised loss as in Eq.~\eqref{eq:loss} using AdamW \cite{loshchilov2019adamw} with a learning rate of $10^{-3}$ for 800 epochs; no ground-truth labels for $(\alpha_i, \beta_i, \gamma_i)$ are required. At each training iteration, receiver positions are first set to the U-Net initialization. The MLP then alternates between predicting coefficients and updating positions for $N$ steps (set to 16 in our experiments), as shown in \Cref{fig:mlp}. At each step, the MLP reads the current local context vector, predicts $(\alpha_i, \beta_i, \gamma_i)$, and applies Eq.~\eqref{eq:update}.
After the final step, the loss in Eq.~\eqref{eq:loss} is evaluated and back-propagated through all $N$ steps to update the MLP parameters.

\section{Experimental Results}\label{sec4}
The evaluated circuits are synthesized and placed using the
3 nm PDK \cite{gt3} at 0.7 V.
The technology parameters are shown in \Cref{BEOL_spec}. The benchmark circuits are summarized in \Cref{tab:benchmark}.
We use PrimeTime \cite{primetime} for timing analysis, PowerWurzel \cite{PowerWurzel} for power analysis, and Voltus \cite{voltus} for static IR-drop analysis.
All physical design steps use IC Compiler II~\cite{icc2}, with PDN pitches of 2.310, 3.168, and 3.276\,$\mu$m on BM1--BM3 and $\times$4 buffers as mesh drivers.

\begin{table}[t]
\footnotesize
\caption{Technology parameters for BEOL and nTSVs}
\label{BEOL_spec}
\renewcommand{\arraystretch}{0.9}
\begin{tabular*}{\columnwidth}{@{\extracolsep{\fill}}lccccc}
\toprule
Layer & \makecell{Width \\ {[nm]}} & \makecell{Pitch \\ {[nm]}} & \makecell{Thickness \\ {[nm]}} & \makecell{Resistance \\ {[$\Omega$/$\mu$m]}} & k \\
\midrule
M4  & 21  & 42  & 42  & 167.0 & 3.9 \\
M3  & 14  & 28  & 28  & 437.5 & 3.9 \\
M2  & 12  & 24  & 24  & 621.8 & 3.9 \\
M1  & 14  & 28  & 28  & 437.5 & 3.9 \\
M0  & 12  & 24  & 24  & 621.8 & 3.9 \\
BPR & 32  & 144 & 90  & 24.3  & 3.9 \\
BM1 & 120 & 231 & 50  & 3.8   & 2.2 \\
BM2 & 160 & 288 & 100 & 1.4   & 2.2 \\
BM3 & 160 & 273 & 100 & 1.4   & 2.2 \\
\midrule
nTSV & \multicolumn{2}{c}{Power-nTSV} & \multicolumn{3}{c}{Clock-nTSV} \\
\midrule
Connectivity & \multicolumn{2}{c}{BM1 and BPR} & \multicolumn{3}{c}{BM1 and M2} \\
Resistance {[$\Omega$]} & \multicolumn{2}{c}{31.3} & \multicolumn{3}{c}{10.8} \\
Dimensions (W/H/T) {[nm]}  & \multicolumn{2}{c}{64 / 32 / 325} & \multicolumn{3}{c}{120 / 96 / 630} \\
KOZ {[nm]} & \multicolumn{2}{c}{200} & \multicolumn{3}{c}{200} \\
\bottomrule
\end{tabular*}
\end{table}

\begin{table}[t]
\centering
\footnotesize
\renewcommand{\arraystretch}{0.95}
\caption{Summary of benchmark designs}
\label{tab:benchmark}
\begin{tabular*}{\columnwidth}{@{\extracolsep{\fill}}lccccc}
\toprule
Circuit & \#Instances & \#Sinks & \makecell{\#Mesh \\ drivers} & \makecell{Area \\ {[$\mu$m$^2$]}} & \makecell{Target freq. \\ {[GHz]}} \\
\midrule
tv80 \cite{opencores}  & 3,530     & 359    & 4    & 275      & 0.78 \\
AES \cite{opencores}   & 11,108    & 530    & 9    & 665      & 1.03 \\
SHA3 \cite{opencores}  & 18,500    & 2,245  & 16   & 1,537    & 1.07 \\
NVDLA* \cite{nvdla_hw} & 1,325,650 & 72,829 & 1,122 & 105,850  & 0.50 \\
\bottomrule
\end{tabular*}
\raggedright\scriptsize *Partition m of NVDLA.
\end{table}

\subsection{Selection of BM Layers for Mesh}
\begin{table}[t]
\centering
\caption{Capacitance of clock mesh straps on candidate layers}
\label{cap_table}
\footnotesize
\renewcommand{\arraystretch}{0.95}
\begin{tabular*}{\columnwidth}{@{\extracolsep{\fill}}lccccc}
\toprule
Layer & M4 & M3 & BM1 & BM2 & BM3 \\
\midrule
Strap width [nm]             & 160   & 160   & 120   & 160   & 160   \\
Strap spacing [nm]           & 9,344  & 9,668  & 1,035  & 9,344  & 9,668  \\
Dist.\ to gnd [nm]    & 413   & 357   & 320   & 1070  & 1870  \\
\midrule
\multicolumn{6}{@{}l}{Capacitance [aF/$\mu$m]} \\
$C_{\text{gnd}}$       & 37.1 & 42.3 & 39.3 & 28.6 & 22.2 \\
$C_{\text{coup}}$      & 5.7  & 6.7  & 4.7  & 5.5  & 4.5  \\
$C_{\text{total}}$     & 42.7 & 48.9 & 44.0 & 34.1 & 26.7 \\
\bottomrule
\end{tabular*}
\end{table}

\begin{table*}[t]
\caption{Performance comparison on four circuits}
\label{tab:power_perf}
\centering
\footnotesize
\setlength{\tabcolsep}{1.2pt}
\renewcommand{\arraystretch}{1.05}
\begin{tabular*}{\textwidth}{@{\extracolsep{\fill}} l cccc | cccc | cccc | cccc @{}}
\toprule
 & \multicolumn{4}{c|}{\textbf{tv80}} & \multicolumn{4}{c|}{\textbf{AES}} & \multicolumn{4}{c|}{\textbf{SHA3}} & \multicolumn{4}{c}{\textbf{NVDLA}} \\
\cmidrule(lr){2-5} \cmidrule(lr){6-9} \cmidrule(lr){10-13} \cmidrule(lr){14-17}
Parameter &
\makecell{FS mesh\\+\cite{tap_driver}} & \makecell{BS mesh\\+\cite{tap_driver}} & \makecell{BS mesh\\+\cite{pfm}} & \makecell{BS mesh\\+Prop.} &
\makecell{FS mesh\\+\cite{tap_driver}} & \makecell{BS mesh\\+\cite{tap_driver}} & \makecell{BS mesh\\+\cite{pfm}} & \makecell{BS mesh\\+Prop.} &
\makecell{FS mesh\\+\cite{tap_driver}} & \makecell{BS mesh\\+\cite{tap_driver}} & \makecell{BS mesh\\+\cite{pfm}} & \makecell{BS mesh\\+Prop.} &
\makecell{FS mesh\\+\cite{tap_driver}} & \makecell{BS mesh\\+\cite{tap_driver}} & \makecell{BS mesh\\+\cite{pfm}} & \makecell{BS mesh\\+Prop.} \\
\midrule
Max local clock latency\textsuperscript{*} [ps]
 & 51.1 & 50.5 & 40.9 & 43.4
 & 40.3 & 38.8 & 36.3 & 33.0
 & 65.3 & 59.2 & 59.1 & 47.3
 & 180.8 & 109.0 & 137.3 & 100.6 \\

Max clock latency [ps]
 & 112.7 & 106.2 & 98.0 & 100.6
 & 158.1 & 155.7 & 150.6 & 151.5
 & 213.7 & 202.8 & 203.5 & 189.2
 & 2070.7 & 1840.3 & 1849.2 & 1684.9 \\

Global skew [ps]
 & 33.2 & 34.5 & 19.9 & 21.1
 & 23.1 & 23.3 & 19.8 & 18.0
 & 48.5 & 42.6 & 39.3 & 24.3
 & 445.4 & 231.0 & 265.1 & 209.7 \\

Max setup skew [ps]
 & 26.0 & 28.2 & 19.9 & 18.2
 & 17.5 & 18.2 & 16.9 & 14.8
 & 41.0 & 38.8 & 38.3 & 23.3
 & 408.0 & 175.9 & 114.8 & 155.6 \\

Worst slack [ps]
 & +176.3 & +189.3 & +206.8 & +185.7
 & +183.2 & +169.1 & +176.5 & +152.1
 & -8.0 & +4.6 & -13.8 & +26.1
 & -227.5 & -238.1 & -212.7 & -215.7 \\
\midrule

Max IR-drop [mV]
 & 2.1 & 3.2 & 4.7 & 1.7
 & 6.7 & 7.9 & 7.9 & 6.6
 & 9.8 & 11.1 & 11.7 & 9.4
 & 38.2 & 41.8 & 56.8 & 38.1 \\

\# KOZ violations
 & 0 & 30 & 0 & 0
 & 0 & 112 & 0 & 0
 & 0 & 240 & 0 & 0
 & 0 & 18768 & 0 & 0 \\
\midrule

Mesh net switching [$\mu$W]
 & 11 & 9 & 9 & 9
 & 44 & 34 & 34 & 33
 & 104 & 82 & 81 & 81
 & 3561 & 2886 & 2847 & 1774 \\

Total clock power [$\mu$W]
 & 352 & 350 & 346 & 348
 & 686 & 676 & 671 & 672
 & 3078 & 3053 & 3042 & 3027
 & 48928 & 48510 & 48239 & 47929 \\
\midrule

\# PFM iterations
 & -- & -- & 72 & 31
 & -- & -- & 256 & 14
 & -- & -- & 76 & 7
 & -- & -- & 1000 & 18 \\
\bottomrule
\end{tabular*}
\begin{flushleft}
*Maximum delay from mesh receiver input to any sink, including mesh receiver cell delay and local wire delay.
\end{flushleft}
\end{table*}

In addition to lower resistance, the BM layers chosen for mesh should also exhibit lower capacitance to take full advantage of backside clock mesh. To support the selection, Raphael \cite{raphael} is applied to accurately extract the capacitances of BM1, BM2, and BM3, as well as those of M3 and M4 in frontside for comparison.
As shown in \Cref{cap_table}, BM2 and BM3 are good choices for clock mesh due to their significantly smaller total capacitance, in comparison to M3 and M4, respectively; this is primarily due to the increased distance to the substrate. BM1 exhibits a total capacitance of 44.0~aF/$\mu$m, comparable to frontside metals due to its close distance to substrate, and can be assumed for local connections between mesh and clock buffers. Clock mesh straps are placed between the PDN straps (see \Cref{cap_table} for width and spacing), and mesh drivers at their intersections are legalized to avoid KOZ violations with power-nTSVs.

\subsection{CDN Quality Assessment}
Four placement options are compared: \cite{tap_driver} applied to FS mesh, \cite{tap_driver} applied to BS mesh, \cite{pfm} applied to BS mesh, and the proposed method on BS mesh. Note that \cite{tap_driver} places one mesh receiver at each uniform grid center without considering KOZ, while the standard PFM \cite{pfm} handles KOZ but relies on globally shared $(\alpha, \beta, \gamma)$. To ensure a competitive baseline, we run \cite{pfm} from the grid-based initialization of \cite{tap_driver}, not random positions, and grid-search $(\alpha, \beta, \gamma)$ over $\{0.25, 0.5, 0.75, 1.0\}$, the same range the MLP can produce. Among the $4^3=64$ combinations, the one minimizing the hard version (exact min/max) of Eq.~\eqref{eq:loss} is selected.

Each benchmark is an independent optimization instance: design-specific U-Net and MLP are trained and applied per design, giving label-free, per-design optimization. Training takes 2.0 to 47.5\,min and actual mesh receiver placement (U-Net inference and MLP-guided PFM refinement) takes 0.04 to 144\,s across the four designs. Standard PFM requires a 64-combination sweep (4.1\,h on NVDLA) to find good weights, whereas our method predicts them directly; including training, our entire flow takes about 50\,min.

\subsubsection{Latency and Skew}
Table~\ref{tab:power_perf} summarizes the results. The proposed method reduces max local clock latency by 15--44\% over FS mesh~+~\cite{tap_driver}. Most of this improvement comes from the ML-guided placement: backside migration alone (FS mesh~+~\cite{tap_driver} vs.\ BS mesh~+~\cite{tap_driver}) contributes only 1--9\% on tv80, AES, and SHA3, and 40\% on NVDLA, whose large mesh has the highest RC. As the local clock latency is the primary source of latency imbalance across sinks, this reduction also lowers global skew\footnote{Max minus min clock latency over all sinks} by 22--53\% and max setup skew\footnote{The worst-case launch-to-capture clock latency difference among all timing paths} by 15--62\%. Since mesh-receiver placement is performed before signal routing, global skew is the directly optimized metric; max setup skew is an indirect benefit on actual timing paths.
Against BS mesh~+~\cite{pfm}, the proposed method loses only 2.5\,ps of max local clock latency on tv80 but achieves 2.8$\times$ lower IR-drop, and is better on the larger designs in both latency and global skew.

\subsubsection{Clock Latency Decomposition}
\begin{figure}[t]
  \centering \includegraphics[scale = 0.98]{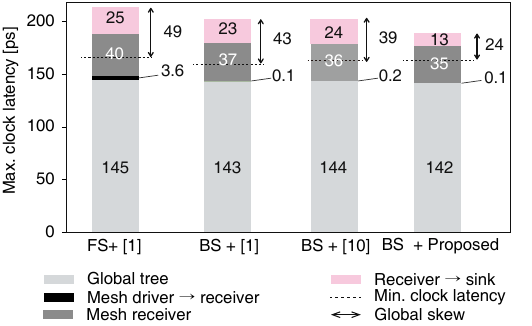}
  \caption{Max clock latency breakdown in SHA3 benchmark.}
  \label{fig:latency_breakdown}
\end{figure}
To identify the source of improvement in latency and skew, \Cref{fig:latency_breakdown} decomposes the max clock latency of SHA3, whose reduction is closest to the four-design average. 
The first two bars, using the same placement method, isolate the effect of migrating the clock mesh from the frontside to the backside. The last three bars compare the placement methods. 
Backside migration leaves the global-tree delay nearly unchanged but reduces the mesh-wire and receiver-cell delays through lower mesh RC and improved input slew. Among the backside cases, the proposed placement further reduces the worst receiver-to-sink delay from 23--24 to 13\,ps by shortening the maximum sink-to-receiver distance. The minimum latency remains comparable while the maximum latency decreases, reducing global skew from 39--43 to 24\,ps.

\subsubsection{Worst Slack}
Worst slack is not a direct objective, as it also depends on datapath delay.
Still, the proposed method performs well: in SHA3, it attains the largest margin (+26.1\,ps) while standard PFM falls
back to -13.8\,ps, and in NVDLA it recovers 22.4\,ps over BS
mesh~+~\cite{tap_driver} and is within 3\,ps of the best.
In tv80 and AES it falls 21.1\,ps and 24.4\,ps below BS mesh~+~\cite{pfm},
but margins of 185.7\,ps and 152.1\,ps remain, which does not affect timing
closure.

\subsubsection{IR-drop}
\begin{figure}[t]
  \centering \includegraphics[scale = 0.95]{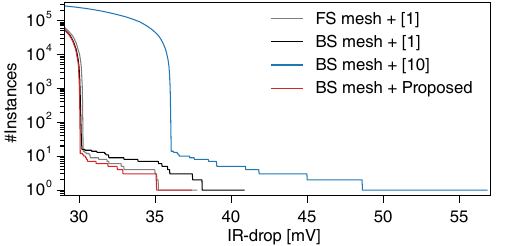}
  \caption{IR-drop distributions on NVDLA: the number of instances whose IR-drop exceeds each value.}
  \vspace{-4mm}
  \label{fig:IR-drop_histogram}
\end{figure}

As shown in \Cref{tab:power_perf}, BS mesh~+~Proposed achieves the lowest max IR-drop. Each baseline degrades IR-drop for a different reason. BS mesh~+~\cite{tap_driver} ignores KOZs, so up to 6\% of power-nTSVs are removed to resolve violations; the proposed method lowers max IR-drop by 9--47\% against it. BS mesh~+~\cite{pfm} includes the same IR-drop-aware repulsive force, but its globally shared weight provides insufficient repulsion near hotspots; on NVDLA, this shifts the distribution upward to 56.8\,mV (\Cref{fig:IR-drop_histogram}). Even against FS mesh~+~\cite{tap_driver}, which has no KOZ issue, the proposed method is 0--19\% lower, as the repulsive force pushes receivers away from high-current regions while preserving all power-nTSVs. 

\subsubsection{Clock Power}
Because BM layers lower the mesh wire capacitance, BS mesh~+~Proposed reduces mesh-net switching power by 18--50\% over FS mesh~+~\cite{tap_driver}; the largest saving is on NVDLA, whose mesh is the largest (3561 to 1774\,$\mu$W).

\subsubsection{PFM Convergence}
The ML-guided PFM converges in 7--31 iterations across all designs, whereas the standard PFM requires 72--256 iterations and, on NVDLA, fails to converge even after 1000 iterations; NVDLA entries in \Cref{tab:power_perf} are the values at this cap. On average, iterations are reduced by 85\%.

\subsection{Ablation Study}
\Cref{tab:ablation} isolates the contribution of each stage on SHA3.
The standard PFM here uses ($\alpha, \beta, \gamma$) set to the mean of the MLP predictions, not the values of \Cref{tab:power_perf}.
The first three rows isolate the MLP contribution under random initialization: standard PFM requires 258 iterations at 232.0\,ps latency and 71.2\,ps skew, while ML-guided PFM converges in 13 iterations at 195.7\,ps and 37.3\,ps.
To confirm the gain is from local context, we invert the MLP predictions, assigning the smallest $\alpha_i$ to the receiver needing the largest.
This variant is worst in both metrics, showing that position-dependent coefficients help only when they capture
local context correctly.

The last four rows start from the U-Net initialization. U-Net initialization alone reaches 192.2\,ps---already better than standard PFM with random init---but leaves 5 KOZ violations, as the U-Net only penalizes them through a loss term.
Standard PFM removes all violations via explicit repulsion at a slight latency cost (192.8\,ps), and ML-guided PFM converges in 7 iterations with the best latency (189.2\,ps), zero violations, and the lowest IR-drop (9.4\,mV).

\begin{table}[t]
\caption{Ablation study on SHA3}
\label{tab:ablation}
\centering
\renewcommand{\arraystretch}{0.95}
\resizebox{\linewidth}{!}{%
\begin{tabular}{ll ccccc}
\toprule
Initialization & PFM method &
\makecell{\#PFM\\iter.} &
\makecell{Max clk \\latency {[ps]}} &
\makecell{Max setup\\skew {[ps]}} &
\makecell{KOZ\\viol.} &
\makecell{Max IR-drop\\{[mV]}} \\
\midrule
Random & Standard            & 258 & 232.0 & 71.2 & 0 & 11.8 \\
Random & ML-guided        & 13  & 195.7 & 37.3 & 0 &  11.6   \\
Random &  ML-guided (inv.)     & 15   & 256.0 & 90.9 & 0 &   12.2  \\
\midrule
U-Net  & None           & --  & 192.2 & 23.7 & 5 & 10.5 \\
U-Net  & Standard            & 18  & 192.8 & 24.0 & 0 & 9.4 \\
\textbf{U-Net}  &\textbf{ ML-guided }       & \textbf{7}   & \textbf{189.2} & \textbf{23.3} &\textbf{ 0} & \textbf{9.4} \\
U-Net  & ML-guided (inv.)      & 7  & 194.0  & 23.0 & 0 &  11.3 \\
\bottomrule
\end{tabular}%
}
\vspace{-3mm}
\end{table}

\section{Conclusion}\label{sec5}
This paper has proposed a CDN architecture with clock mesh on BM layers. The KOZ between power-nTSV and clock-nTSV should be respected because BM layers are shared between CDN and PDN. This can be addressed by careful placement of mesh receivers, which we have addressed in this paper. A U-Net predicts the initial locations of mesh receivers and an MLP predicts position-dependent force weights for PFM. Both models are trained via self-supervised learning using a differentiable objective that optimizes clock latency, IR-drop, and KOZ violations.
Experiments have demonstrated the benefit of the proposed CDN: 26\% lower maximum local clock latency, 40\% smaller global skew, and 29\% less mesh switching power on average than the frontside baseline \cite{tap_driver}. Compared with standard PFM on the same backside CDN, the ML-guided PFM has eliminated all KOZ violations with 33\% lower IR-drop and 85\% fewer iterations, without any manual weight tuning.

\bibliographystyle{ACM-Reference-Format}
\bibliography{ref}
\end{document}